\documentclass[%
 reprint,
 superscriptaddress,
 bibnotes,
 amsmath,amssymb,
 aps,
prl,
]{revtex4-2}

\usepackage{tikz}

\usepackage[margin=2.5cm]{geometry}

\newcommand{\textcircledcor}[1]{\raisebox{.5pt}{\textcircled{\raisebox{-.9pt} {#1}}}}
\usepackage{amsmath}

\usepackage[section]{placeins}
\usepackage{gensymb}

\usepackage{graphicx}
\graphicspath{{figures/},{appendix_figures/}}
\usepackage{appendix}

\usepackage{verbatim}
\usepackage{dcolumn}
\usepackage{bm}

\usepackage{color}

\begin{document}


\title{What is the stiffness of a bent book?}%

\author{Samuel Poincloux}
\author{Tim Chen}%
\affiliation{Flexible Structures Laboratory, \'Ecole Polytechnique F\'ed\'erale de Lausanne (EPFL), Lausanne, Switzerland}%

\author{Basile Audoly}
\affiliation{Laboratoire de M\'ecanique des Solides, CNRS, Institut Polytechnique de Paris, France}%
\author{Pedro Reis}
\affiliation{Flexible Structures Laboratory, \'Ecole Polytechnique F\'ed\'erale de Lausanne (EPFL), Lausanne, Switzerland}%

\date{\today}
\begin{abstract}
We study the bending of a book-like system, comprising a  stack of elastic plates coupled through friction. The behavior of this layered system is rich and nontrivial, with a non-additive enhancement of the apparent stiffness and a significant hysteretic response.  A dimension reduction procedure is employed to develop a centerline-based theory describing the stack as a non-linear planar rod with internal shear. We consider the coupling between the nonlinear geometry and the elasticity of the stacked  plates, treating the interlayer friction perturbatively. This model yields predictions for the stack's mechanical response in three-point bending that are in excellent agreement with our experiments. 
Remarkably, we find that the energy dissipated during deformation can be rationalized over three orders of magnitude, including the regimes of a thick stack with large deflection. This robust dissipative mechanism could be harnessed to design new classes of low-cost and efficient damping devices.
\end{abstract}

\maketitle

Multilayered microstructure layouts are essential in many biological and engineered materials for enhanced mechanical properties~\cite{vinson_sandwich_2001}. For example, nacre and nacre-like materials have been investigated and engineered for their superior stiffness, strength, and toughness~\cite{jackson1988mechanical,tang2003nanostructured,yin2019impact}. Layered architectures are also found across scales, from multilayer graphene~\cite{wang_bending_2019} and fish scales~\cite{bruet2008materials,ali2019frictional}, to deployable mechanisms~\cite{umali_vibration_2017} and geological stacks~\cite{ran_shear_1994}. In all these systems, interlayer interactions dictate the overall mechanical response. Frictional damping across layered elements is also central to the performance of classic engineering systems such as mechanical joints~\cite{bograd2011modeling}, turbine blades~\cite{griffin_1990}
and leaf springs~\cite{badrakhan_slip_1994,osipenko_contact_2003}. 
There has been progress in modeling layered system with a few number of interfaces~\cite{hansen_structural_1997,sedighi_stick-slip_2013,asker_dynamic_2018} or when frictional effects dominate~\cite{alarcon_self-amplification_2016}. Still, it remains challenging to predict how the microscopic architecture and interlayer interactions of a layered mechanical system give rise to a specific macroscopic constitutive response, especially for large deformations.

\begin{figure}[b]
\includegraphics[width=\columnwidth]{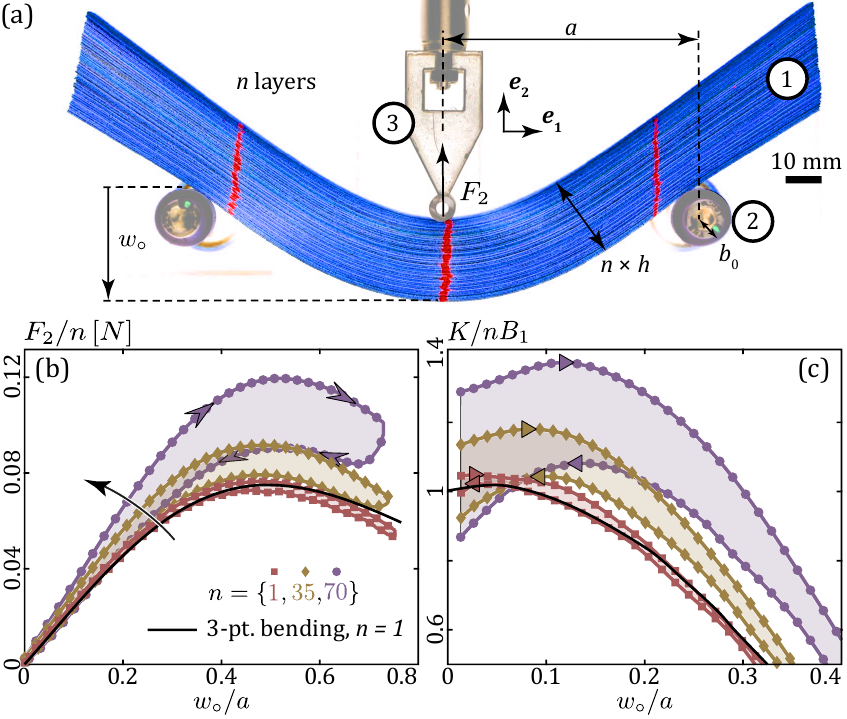}
\caption{\label{fig:Fig1}
(a)~Photograph (front view) of the experimental setup. A stack~$\textcircledcor{1}$ of $n$ plates ($n=70$ here) is placed on two rollers~$\textcircledcor{2}$ and loaded by an indenter~$\textcircledcor{3}$ prescribing the deflection at mid-span.  (b)~Loading-unloading curves of the average intender force per plate, $F_2(w_\circ)/n$, for selected values of $n$. The thin black line corresponds to the classic, nonlinear prediction for the 3-point bending of a single plate, $n=1$~\cite[S.II]{SupInf_PRL}. (c)~ Normalized incremental rigidity, $K/(nB_1)$, and its maxima, $K_\mathrm{m}^{\pm}$ ($\rhd$ and $\lhd$ symbols for loading and unloading, respectively). 
}
\end{figure}

Here, we study the mechanics of a model layered system, where the effects of the small-scale structural layout and friction can be related directly to the macroscopic response. Specifically, we address the question: \textit{What is the stiffness of a bent book?} It is well known that the bending stiffness of a slender structure scales as its thickness cubed, $\sim h^3$~\cite{Audoly:2010Book}. Naturally, the answer for a book with $n$ sheets is bound by the two limiting cases of $\sim (nh)^3$ and $\sim nh^3$. The first estimate ignores the possibility of sliding (infinite friction), whereas the second neglecting the interlayer shear stresses (zero friction). Computing the correct answer for finite friction is nontrivial due to nonlinear coupling between elasticity, nonlinear geometry and friction in this non-conservative layered system. We study this problem by performing precision nonlinear bending tests of a multi-layered stack of elastic plates interacting solely through friction (see Fig.~\ref{fig:Fig1}a). We quantify the mechanical response of this book-like system, including the dissipated energy. Following a dimension reduction procedure, we develop a beam-like theory based on the centerline of the stack. This model takes into account the nonlinear geometry of large stacks and treats friction as a perturbation. 

In our experiments, we quantify the resistance to bending of a book-like system by performing mechanical tests of a stack of $n$ plates in a 3-point bending configuration (see photographs of the apparatus in Fig.~\ref{fig:Fig1}a and ~\cite[S.I]{SupInf_PRL}). 
This canonical testing geometry is well-established for the characterization of the mechanics of beams, including in the large deflection regime~\cite{ohtsuki_analysis_1986,batista_large_2015}. We seek to quantify the effect of frictional dissipation between the plates on the mechanical response of the system. Our stack comprises $n$ plates made of PolyEthylene Terephthalate (PET, Partwell group), each with dimensions $2L\times W\times h=220\times 30 \times 0.286\,\mathrm{mm}^3$. The number of plates is varied in the range $1\leq n \leq 70$. Both faces of the plates are roughened using sandpaper (K$80$, Emil-Lux Gmbh) to avoid interlayer adhesion and ensure reproducible dry-friction interactions~\cite{baumberger2006solid}. The 3-point bending configuration is established by two fixed lower supports, separated by $2a=130\,\mathrm{mm}$, and an indenter located at mid-span. The fixed supports are set as rollers, comprising two steel cylinders (radius $b_0=6.8\,\mathrm{mm}$) coated with a film of VinylPolySiloxane (thickness $\approx100\,\mu\mathrm{m}$) to prevent sliding, and mounted on air-bearings (IBS Precision Engineering, pressure $\approx70\,\mathrm{psi}$) to offer nearly frictionless rotation. The reaction force at the indenter, $F_2$, is measured by a universal testing machine (Instron 5943). The imposed-displacement indentation is performed cyclically, at constant speed ($v=\pm1\,\mathrm{mm/s}$), such that the mid-span deflection is varied in the range $0\leq w_\circ \leq w_\circ^{\text{max}}$. The geometry and loading conditions ensure that each plate remains in the elastic regime. Our experimental apparatus yields highly reproducible and precise mechanical response measurements (further evidence provided in \cite[S.I]{SupInf_PRL}).

In Fig.~\ref{fig:Fig1}(b), we plot representative curves of the average load per plate, $F_2/n$, for cycles with amplitude $w_\circ^{\text{max}}=50\,\mathrm{mm}$, at selected values of $n=\{1,\, 25,\, 70\}$. For $n=1$, the response agrees with the classic prediction for  large-deflection 3-point bending; there is a linear regime followed by a maximal load with no hysteresis during unloading (see~\cite[S.II]{SupInf_PRL}). 
We find that both the maximal load per layer $F_2/n$ and the energy dissipation through friction (area of the hysteresis loop) increase with $n$, implying that the behavior of the stack is not a superposition of $n$ independent layers.
To address this nonlinear response, we introduce the incremental stiffness 
$K(w_\circ)=\frac{a^3}{6}\frac{\mathrm{d}F_2}{\mathrm{d}w_\circ}$; the prefactor ensures that $K= nB_1$ for small deflection and without friction, where $B_1$ is the bending rigidity of a single plate, $B_1=\frac{Eh^3W}{12(1-\nu^2)}$, $E$ is the Young modulus and $\nu$ is the Poisson's ratio.
In Fig.~\ref{fig:Fig1}(c), we plot $K(w_\circ)/(nB_1)$, using the same data as in Fig.~\ref{fig:Fig1}(b), for a loading-unloading cycle. 
The limiting value $K(0)/(nB_1)$ for small deflections is 1 for $n=1$, and increases with $n$, implying that, when $n>1$, friction affects even the initial response.
In addition, the loading curves display an increasingly pronounced hysteresis as $n$ increases: the incremental stiffness $K$ is different between loading and unloading. The maximum stiffness, $K_\mathrm{m}^\pm$, provides a robust measure of the bending rigidity of the stack; we define one for loading, $K_\mathrm{m}^+$, and one for unloading, $K_\mathrm{m}^-$.

\begin{figure}[b]
\includegraphics[width=\columnwidth]{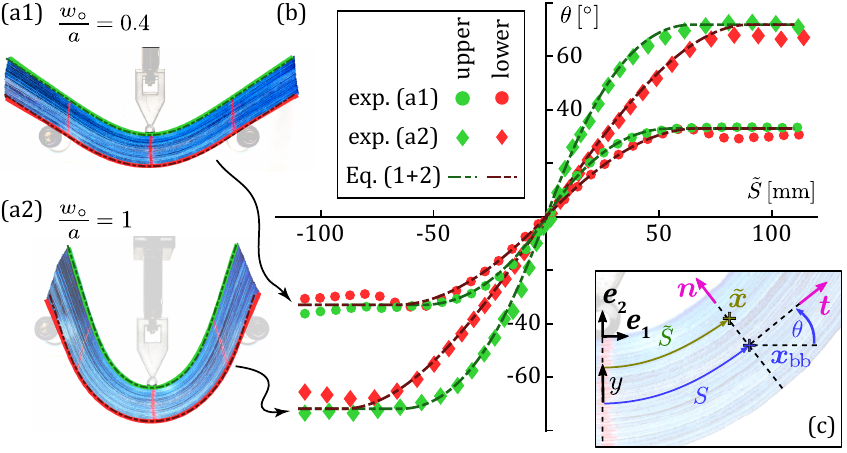}
\caption{\label{fig:Fig2} (a) Snapshots of a deformed stack with $n=70$, at (a1) $w_\circ/a=0.4$ ($\circ$) and (a2) $w_\circ/a=1$ ($\diamond$). (b)~Schematic diagram of the geometric quantities used in~Eq.~(\ref{Eq:geom_stack}).
(c)~Tangent angle $\theta$ versus arc length $\tilde{S}$ along the uppermost and lowermost plates (open and filled symbols, respectively). The predictions (dashed lines) were obtained by integrating Eq.~(\ref{Eq:elastica}). The predicted profiles are superimposed as colored curves in (a1) and (a2).
}
\end{figure}

Having characterized the overall loading response of our stacks, we proceed by further quantifying the kinematics of a bent stack. By way of example, we select a thick stack with $n=70$ and focus on two representative configurations at moderate and large deflections, $w_\circ/a=0.4$ and $1$, see Fig.~\ref{fig:Fig2}(a1,a2). The schematic diagram in Fig.~\ref{fig:Fig2}(c) defines the quantities used in the geometric analysis:
$\theta$ is the tangent orientation, and $\tilde{S}$ is the arc length measured along a specific plate. In Fig.~\ref{fig:Fig2}(b), we plot the profiles $\theta(\tilde{S})$ for the uppermost and lowermost plates, for the two selected indentation levels. The profiles of the uppermost and lowermost plates are different, especially at larger indentations: the former is more localized than the latter. An appropriate model for the stack must account for these through-thickness variations.

We visualize the extent of shear in our book-like system by physically painting three red lines on the lateral face of the stack (see Fig.~\ref{fig:Fig2}a1,a2), perpendicularly to its centerline in the undeformed configuration. During the deformation ensued by the 3-point bending test, we find that the two outer lines lose perpendicularity to the centerline, indicating that there is significant shear. The non-penetration of the contacting plates is at the source of this shear build-up, which is known to arise in parallel bundles of inextensible curves~\cite{pham1992offset}; strong geometric constraints couple the layers in the stack.
A well-known model for thick and shearable beams is that of Timoshenko~\cite{timoshenko_correctionfor_1921,li_large_2016}; however, it is inapplicable here as it assumes that the shear stress has an elastic origin.

To rationalize our experimental results, we build a 1D model for thick beams that accounts for internal friction at the interfaces of the layers. Similar reduction methods for bundles of slender components have recently been employed to describe helical strips~\cite{ansell_threading_2019,noselli_smart_2019} or bundled filaments~\cite{atkinson_constant_2019}, albeit in different geometries than ours and without considering friction. The centerline of the stack is represented as an inextensible curve $\mathbf{x}_\mathrm{bb}(S)$ with arc length $S$ and curvature $\kappa(S)$; we reserve the symbol $S$ for arc lengths measured along the stack's centerline, whereas $\tilde{S}$ pertains to the arc length along a specific plate.
The transverse coordinate $y$ varies from $-nh/2$ at the lowermost plate to $nh/2$ at the uppermost one. In the absence of delamination, the final position of a point belonging to the plate offset by $y$ from the stack's centerline writes as
\begin{equation}
\tilde{\mathbf{x}}(S,y)=\mathbf{x}_\mathrm{bb}(S)+\mathbf{n}(S)y,
\label{Eq:geom_stack}
\end{equation}
where $\mathbf{n}(S)$ is the unit normal to the centerline (Fig.~\ref{fig:Fig2}c).
Note that in our non-Lagrangian parameterization, the final position $\tilde{\mathbf{x}}$ is viewed as a function of the arc length $S$ of its \emph{projection} $\mathbf{x}_\mathrm{bb}$ onto the centerline in the \emph{final} configuration. Thus, $S$ is different from the Lagrangian arc length $\tilde{S}$, and $\tilde S(S,y)-S$ provides a measure of shear.
The two arc lengths are related as $\mathrm{d}\tilde S=(1-y\kappa(S))\mathrm{d}S$
due to the combined effects of curvature and plate inextensibility, as shown in~\cite[S.II]{SupInf_PRL} by differentiating Eq.~(\ref{Eq:geom_stack}).

From~Eq.~(\ref{Eq:geom_stack}),  the curvature of a plate is $\tilde{\kappa}(\tilde S,y) = \kappa(S)(1 - y \kappa(S))^{-1}$. The bending energy $\mathcal{E}$ of the stack is found by summing the contributions $\int_{-L}^{+L}\frac{B_0}{2}\tilde{\kappa}^2\mathrm{d}\tilde{S}$ from each plate, yielding 
$\mathcal{E} =2\frac{B_0}{nh^2} \,\int_{0}^\ell \varphi(n h\kappa(S))\,\mathrm{d}S$, where
$\varphi(x)=\frac{x}{2}\,\ln\left(\frac{1+\frac{x}{2}}{1-\frac{x}{2}}\right)$ and $\ell$ is the arc length where contact with the rollers takes place, $\ell(w_\circ=0)=a$ (see~\cite[S.II]{SupInf_PRL}). 
The range of the integration to obtain $\mathcal{E}$ has been restricted to $0\leq S\leq \ell$, given both the symmetry of the solution and the fact that the overhanging parts of the slack beyond the supports remain straight and, therefore, carry no energy.

The strain energy potential $\mathcal{E}$ defines an equivalent nonlinear beam model for the stack, with an internal moment given by the constitutive law $M(S)=\frac{B_0}{h}\,\varphi'(nh\,\kappa)$. Following a variational approach (see~\cite[S.II]{SupInf_PRL}), one obtains the governing equilibrium (Kirchhoff) equations for planar rods,
\begin{equation}
       \frac{nB_1\,\theta'' (S)}{\left({1-{
       \frac{n^2 h^2 \theta'^2(S)}{2}}}\right)^2} +
        \left(\frac{F_2}{2}\cos \theta (S)  + F_1\sin \theta (S) \right)= 0, \label{Eq:elastica}
\end{equation}  
where primes denote differentiation with respect to $S$, $F_2$ is the indentation force,
and the reaction force at the support $S=\pm\ell$ is written as $\mathbf{F}=\mp F_1\mathbf{e_1}+(F_2/2)\mathbf{e_2}$.
The centerline satisfies $x_1'(S)=\cos\theta(S)$ and $x_2'(S)=\sin\theta(S)$.
The boundary conditions are $(\theta,x_1,x_2)_{S=0}=(0,0,-w_\circ)$ and $(\theta',x_1,x_2)_{S=\ell}=(0,a-b\sin \theta(\ell),b(\cos\theta(\ell)-1))$, with $b=b_0+nh/2$ as the effective radius of the support. 

Solving the boundary-value problem in~Eq.~(\ref{Eq:elastica}) yields the centerline $\mathbf{x}_\mathrm{bb}(S)$; this solution ignores friction and will be referred to as the \emph{elastic backbone}.
The shape of the full stack can be reconstructed using~Eq.~(\ref{Eq:geom_stack}). 
In~Fig.~\ref{fig:Fig2}(a), we find excellent agreement between the computed and the experimental shapes of the uppermost and lowermost plates. 
As part of the solution process, one also obtains the indentation force $F_{2,\mathrm{bb}}(w_\circ)$.

Next, we address the interlayer friction to rationalize the hysteresis observed in the experiments. Treating friction as a perturbation, we use the (frictionless) elastic backbone solution obtained above to estimate the power $\mathcal{P}_\mu$ dissipated by friction.
This $\mathcal{P}_\mu$ is the integral over all the plate-plate interfaces of the sliding velocity multiplied by the tangential contact stress. From Amontons-Coulomb law of friction, the tangential contact stress is the friction coefficient $\mu$ times the normal stress $\Sigma(S,y)$.
Reconstructing the stress $\Sigma(S,y)$ in the backbone solution and carrying out a partial integration in the transverse direction, one obtains the expression of the dissipated power as~\cite[S.III]{SupInf_PRL}
\begin{equation}
    \mathcal{P}_\mu=\mu h \,n\, | \mathbf{F} | \, | \dot{\theta} (\ell) |
    + 2\mu\int_{0}^{\ell} Q(S)\, | \dot{\theta} (S) |\,\mathrm{d}S
    , \label{Eq:dissip_nonlin}
\end{equation}
where dots denotes differentiation with respect to time and $Q(S)=\int_{-nh/2}^{nh/2} |\Sigma(S,y)| \mathrm{d}y$.

The first term in~Eq.~(\ref{Eq:dissip_nonlin}) represents the dissipation by the point-like contact force at the supports, while the second term is the dissipation everywhere else in the stack. By symmetry, there is no sliding (hence, no dissipation) at the indentation point. The indentation force is then derived by a global balance of power as 
\begin{equation} 
F_2\,\dot{w_\circ}=-\dot{\mathcal{E}}+\mathcal{P}_\mu \text{.}
\label{Eq:power_balance}
\end{equation}
Whereas $\dot{w_\circ}$ and $\dot{\mathcal{E}}$ change sign between loading and unloading, $\mathcal{P}_\mu$ does not, implying that $F_2$ is different during the two phases.

\begin{figure}[t]
\includegraphics[width=1\columnwidth]{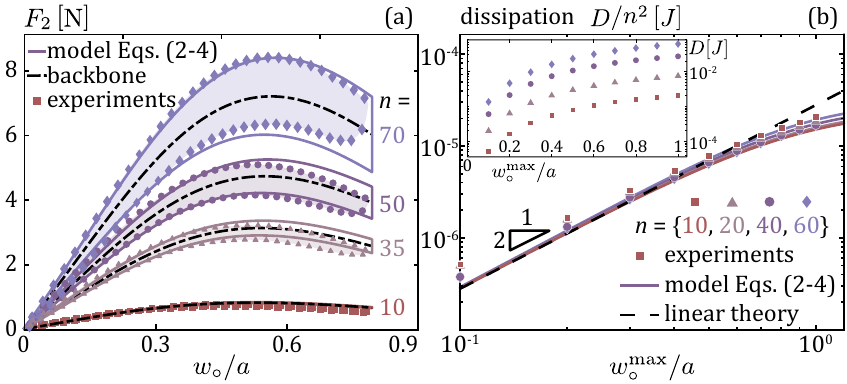}
\caption{\label{fig:Fig3} Predictions of the model from Eqs.~(\ref{Eq:elastica}--\ref{Eq:power_balance}), solid lines, superposed onto the experimental data (points), for stacks with different numbers of layers.
(a)~Load-indentation mechanical response, for loading/unloading cycles. $n=\{ 10,\, 35,\, 50,\, 70\}$. Dashed-dot line corresponds to the elastic backbone (without friction). (b)~Scaled energy dissipated in the stack, $D/n^2$, per cycle, as a function of the scaled indentation amplitude, $w_\circ^{\text{max}}/a$. Inset: Raw data for $D$ versus $w_\circ^{\text{max}}/a$. $n=\{ 10,\, 20,\, 30,\, 50\}$.}
\end{figure}

Before the indentation force can be computed from Eqs.~(\ref{Eq:dissip_nonlin}--\ref{Eq:power_balance}), the kinematic friction coefficient $\mu$ must still be obtained.
Friction coefficients for dry surfaces are known to be sensitive to the magnitude of the normal load~\cite{alarcon_self-amplification_2016}, which, in our system, varies significantly depending on both the amount of indentation and the position along the stack~\cite[S.III]{SupInf_PRL}.
Therefore, an independent measurement of $\mu$ may not be relevant.
Instead, we proceed by extracting $\mu$ directly from the experimental data by leveraging the variations of the stacks stiffness $K_\mathrm{m}^{\pm}$ as a function of $n$. In the limit of small deflections (see~\cite[S.IV]{SupInf_PRL}), our model yields  $K_\mathrm{m}^{\pm} = K_\mathrm{m,bb}\cdot\left(1\pm n\mu\frac{3}{4}\frac{h}{a}\right)$ with $K_\mathrm{m,bb}$ as the bending stiffness of the backbone solution. Exploiting the linear relation between $K_m^{\pm}$ and $n$ provides the friction coefficient as $\mu=0.52\pm0.03$ (see~\cite[S.V]{SupInf_PRL} for more details on how $K_m^{\pm}$ and $\mu$ where obtained from the experimental data). The indentation force $F_2$ can now be obtained from~Eqs.~(\ref{Eq:elastica}--\ref{Eq:power_balance}) to compute the loading curves over the entire indentation range. In Fig.~\ref{fig:Fig3}a, we compare the predictions from our model (solid  curves) with the experiments (data points), finding excellent agreement between the two,
for different values of $n$, with a single parameter $\mu$ that was fitted to the data once and for all.

In Fig.~\ref{fig:Fig3}(b), the energy $D$ dissipated during one loading cycle is plotted as a function of the scaled maximum indentation depth. From the experimental data, $D$ is measured as the area enclosed by the loading-unloading curves. The model predictions are accurate over the entire range of parameters, from thin to thick stacks and small to large deflections; \textit{i.e.},
$10 \leq n\leq 60$ and $ 0.1 \leq w_\circ^{\mathrm{max}}/a \leq 1$, respectively. For small indentations, the curves collapse onto a straight line in the logarithmic plot, corresponding to the power-law $D_\mathrm{lin}=\frac{9\mu B_1 h}{2 a^2}\left({n w_\circ^{\mathrm{max}}/a}\right)^2$ applicable to small deflections (see~\cite[S.IV]{SupInf_PRL}).

In closing, we highlight that the ability of our centerline-based theory to accurately capture the mechanical behavior of a stack of frictional plates was \textit{a priori} not straightforward, given the non-conservative nature of the system. We circumnavigated this challenge by treating friction perturbatively while tracking the localized dissipative regions and considering the full coupling between elasticity and nonlinear geometry. Regarding potential applications, our most significant result is that the energy dissipated per cycle can vary over orders of magnitude. The mechanism that we have uncovered for stacks of frictional plates could be harnessed to design new classes of low-cost and efficient damping devices. As geometry, elasticity, and friction are the sole ingredients, the proposed dissipative mechanism should be applicable across a wide range of length scales. Whereas we focused on a quasi-static setting, dynamic and impact conditions should also be included in future research efforts, which we hope the current study will instigate.

\begin{acknowledgments}
\end{acknowledgments}

\end{document}


\centerline{\bf \large What is the stiffness of a bent book?} 
\vspace{5mm}
\centerline{\bf \large -- Supplemental Information -- }
\vspace{5mm}
\centerline{\bf \large S. Poincloux, T. Chen, B. Audoly, P. Reis}

\author{S. Poincloux, T. Chen, B. Audoly, P. Reis}
\vspace{2cm}

\beginsupplement

\section{Experimental reproducibility}

\begin{figure}[h]
\includegraphics[width=0.7\columnwidth]{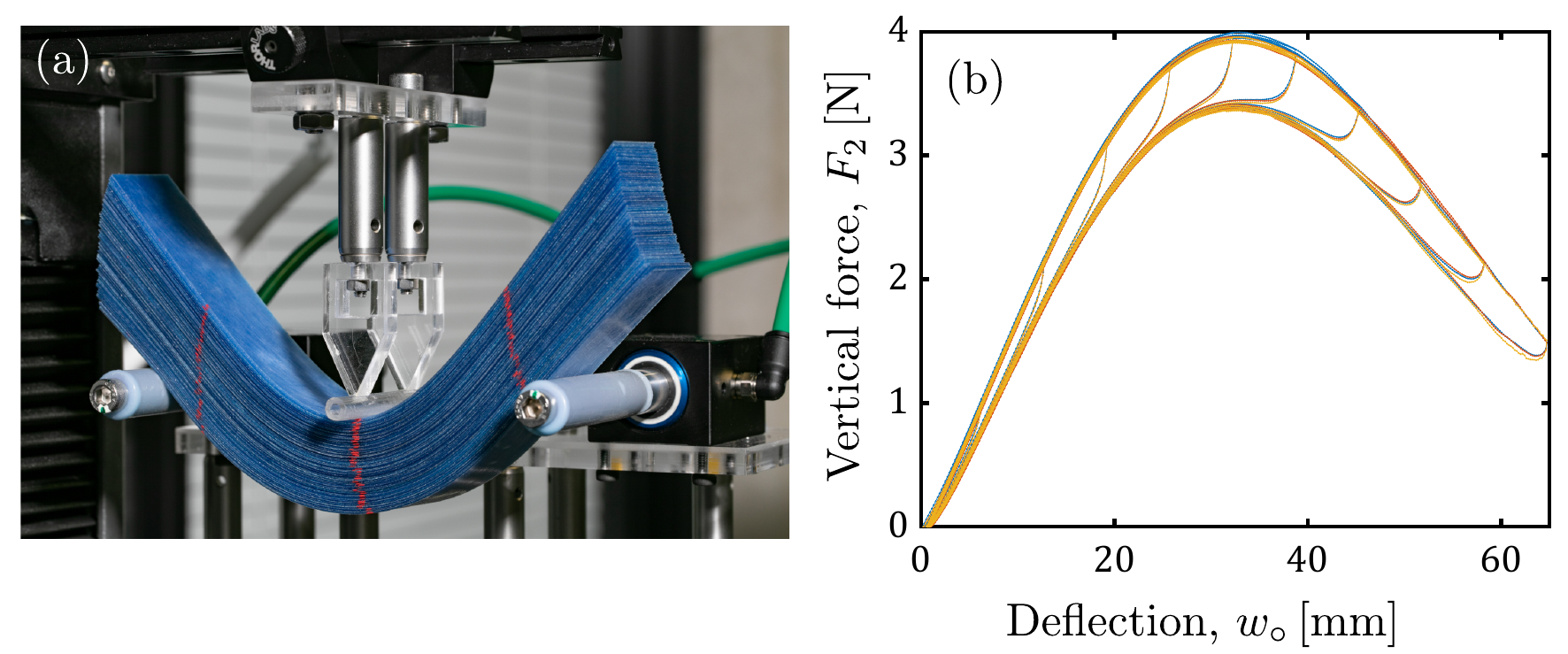}
\caption{\label{fig:FigS1} (a)~Picture of the experimental set-up (photo taken by Alain Herzog). (b)~Vertical force against deflection curve $F_2(w_\circ)$ for $n=40$. At each cycle, $w_\circ^{\mathrm{max}}$ is incrementally increased by $6.5\,\mathrm{mm}$ from $6.5\,\mathrm{mm}$ to $65\,\mathrm{mm}$ ($a=65\,\mathrm{mm}$). The three colors represent three different tests with the stack shuffled in-between. Both loading and unloading curves are highly reproducible. 
}
\end{figure}

In~Fig.~\ref{fig:FigS1}(a), we present a photograph of the full experimental apparatus described in the main text. This set-up provides highly reproducible mechanical response of the stacks upon multiple loading cycles and shuffling of the layers. This reproducibility is highlighted in~Fig.~\ref{fig:FigS1}(b) where we show the raw force-deflection curves $F_2(w_\circ)$ for a stack with $n=40$ plates. The stack is bent cyclically, while gardually incrementing the maximum deflection $w_\circ^{\mathrm{max}}$ within the range $6.5 \leq w_\circ^{\mathrm{max}}\,[\mathrm{mm}] \leq 65$, in steps of $6.5\,\mathrm{mm}$ at each cycle. The experimental test is repeated three times (represented by the three different color in~Fig.~\ref{fig:FigS1}(b), while maintaining the  same number of layers but shuffling them in-between the experiments. For each cycle and for each experimental run, the data overlaps, attesting the reproducibility of our experimental measurements.  

\section{Geometry and energy of the reduced model for the stack}

\begin{figure}
\includegraphics[width=0.6\columnwidth]{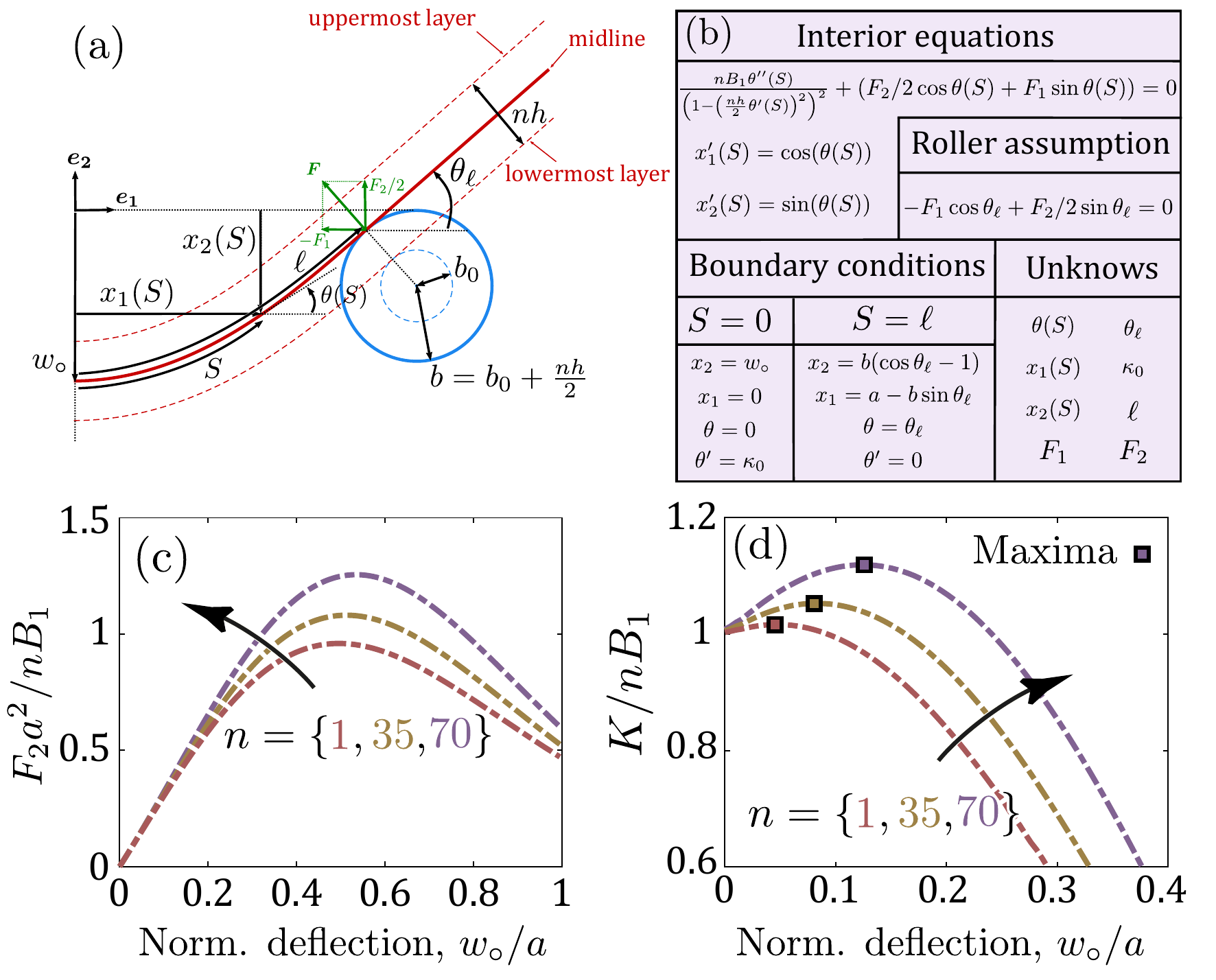}
\caption{\label{fig:FigS2} (a)~Schematic diagram of one  half of a bent stack, illustrating the notations used. The centerline is shown in red (continuous red line) together with the uppermost and lowermost plates (dashed red lines). The physical extent of the roller is shown using a dashed blue circle, while the continuous blue line indicates the effective support interacting with the centerline. (b)~Summary of the equations and variables. (c)~Normalized load-deflection curves computed by solving the  boundary-value problem summarized in~(b) and detailed in the text. (d)~Normalized incremental rigidity computed from the force curves in~(c). 
}
\label{fig:SI:kinematics}
\end{figure}

\noindent \textbf{Kinematics of the stack:} Consider a reference inextensible curve $\mathbf{x}_\mathrm{bb}(S)=x_1(S)\mathbf{e_1}+x_2(S)\mathbf{e_2}$ in the plane, and
its tangent
\begin{equation}
    \mathbf{t} (S) = \frac{\mathd \mathbf{x}}{\mathd S}
\end{equation}
as well as the normal $\mathbf{n} (S)$; see Fig.~2(c) of the main text and Fig.~\ref{fig:SI:kinematics}(a) for schematic diagrams. The curvature is $\kappa=\mathd \theta/\mathd S$ with $\theta$ as the angle between $\mathbf{t}$ and the horizontal axis. The curve with offset $\tilde{\mathbf{x}} (S, y)$ is defined by the non-normal parametrization
\begin{equation}
    S \mapsto \tilde{\mathbf{x}_\mathrm{bb}} (S, y) =\mathbf{x} (S) + y\mathbf{n} (S).
\end{equation}
The arclength $\tilde{S}$ on the offset curve satisfies
\begin{equation}
    \mathd \tilde{S} = | \mathd \tilde{\mathbf{x}_\mathrm{bb}} | = | \mathbf{t} \mathd
   S + y \mathd \mathbf{n} | = (1 - y \kappa) \mathd S.
   \label{Eq:layer_metric}
\end{equation}
We always assume $| y \kappa | < 1$ (no cusp). The offset curves remain parallel to the centerline $\mathbf{x}$. The tangent to the offset curve at $\tilde{\mathbf{x}} (S, y)$ is parallel to $\mathbf{t} (S)$. As a result, the curvature of the offset curve reads
\begin{equation}
    \tilde{\kappa} = 
   \frac{\mathd \theta}{\mathd \tilde{S}} = \frac{\kappa}{\mathd \tilde{S} /
   \mathd S} = \kappa \cdot (1 - y \kappa)^{- 1}.
   \label{Eq:KappaTilde}
\end{equation}

\noindent \textbf{Energy of the stack:} The contribution to the bending energy density $\tilde{\mathcal{E}}$ arising from the `sector' spanned by $\mathd S$ reads
\begin{equation}
    \tilde{\mathcal{E}} \mathd S = \int\limits_{- \frac{nh}{2}}^{+ \frac{nh}{2}} \frac{\mathd y}{h} 
   \frac{B_1}{2}  \tilde{\kappa}^2 \mathd \tilde{S} = \int\limits_{- \frac{nh}{2}}^{+
   \frac{nh}{2}} \frac{\mathd y}{h}  \frac{B_1}{2} \kappa^2 \cdot (1 - y
   \kappa)^{- 1} \mathd S = \frac{B_1}{2 h} \kappa \ln \left(
   \frac{1 + \frac{nh \kappa}{2}}{1 - \frac{nh \kappa}{2}} \right)\mathd S.
\end{equation}   
This can be rewritten as
\begin{equation}
    \tilde{\mathcal{E}} (\kappa) = \frac{B_1}{2h} \kappa \ln \left( \frac{1 + \frac{nh
   \kappa}{2}}{1 - \frac{nh \kappa}{2}} \right) = \frac{B_1}{nh^2} \varphi (nh
   \kappa),
\end{equation}
where
\begin{equation} 
\varphi (k) = \frac{k}{2} \ln \left( \frac{1 + \frac{k}{2}}{1 -
   \frac{k}{2}} \right) 
   \text{.}
   \label{Eq:Aux_fun}
\end{equation}
The Taylor expansion $\varphi (k) = \frac{k^2}{2} + \frac{k^4}{24} + \cdots$ yields
\begin{equation} 
\tilde{\mathcal{E}}  (\kappa) = B_1 n \left( \frac{\kappa^2}{2} + \frac{(nh)^2}{24} \kappa^4 +\cdots \right).
\end{equation} 
The first term in the expression above corresponds to the assumption that
all layers have identical curvature $\kappa$. The subsequent terms bring in a non-linear correction that accounts for the fact that the different layers have different curvatures.\\

\noindent \textbf{Analogous \textit{Elastica} formulation}
\noindent We assume the left/right mirror symmetry is preserved, such that only half of the stack is considered, with $S=0$ as the point of indentation and $S=\ell$ as the contact point with the support, the equilibrium equation is found by requiring that the total energy is stationary. The energy in this problem consists of the bending energy $\mathcal{E}/2=\int_{S=0}^\ell\tilde{\mathcal{E}}\mathd S$. The frictionless contact with the support is enforced through Lagrange multipliers $F_1$ and $F_2$ defined as positive, such that the quantity
$\mathbf{F}=-F_1\mathbf{e_1}+F_2/2\mathbf{e_2}$ can be interpreted as the contact force with the support at $S=\ell$.  Following a variational approach, we arrive at the following equilibrium equation:%
\begin{equation}
       \frac{nB_1\theta'' (S)}{\left({1-\left({\frac{nh}{2}\theta'(S)}\right)^2}\right)^2} +
        \left({F_2/2\cos \theta (S)  + F_1\sin \theta (S) }\right)= 0, \label{Eq:elastica}
\end{equation}  
with the primes denoting differentiation with respect to $S$. The shape of the centerline $\mathbf{x}(S)$ is reconstructed from the solution $\theta(S)$ using $x_1'(S)=\cos(\theta)$ and $x_2'(S)=\sin(\theta)$. The symmetry of the problem imposes the following boundary condition on $S=0$: $\theta(S=0)=0$ and $x_1(S=0)=0$. The displacement is controlled; \textit{i.e.}, $x_2(S=0)=-w_\circ$. The parameter $\ell$ is an unknown whose initial value $\ell=a$ increases as the deflection gets larger. We assume that the layers in the cantilevering portion of the setup extend sufficiently so that $\ell$ is below the physical half-length $L$ of the plates. The part of the stack past the roller ($\ell\leq S\leq L$) does not deform, nor it contributes to the bending energy. At $S=\ell$, the tangent angle $\theta_\ell=\theta(S=\ell)$ is not known; it is effectively a free end and we enforce the boundary condition $\theta'(S=\ell)=0$. The parameter $\ell$ is set by the condition of contact with the support, $x_1(S=\ell)=a-b\sin\theta_\ell$, $x_2(S=\ell)=b(\cos\theta_\ell-1)$, with $b=b_0+\frac{nh}{2}$ being the effective radius of the supports taking the thickness of the stack into account. The section of the plates beyond the contact point in $S=\ell$ follow a continuity condition in position and tangent. 
Finally, as the supports can freely rotate, the reaction force at $S=\ell$ must remain normal to the stack. The support creates a jump of $|\mathbf{F}|$ in the internal normal force of the stack. Being a normal force, it also imposes a relation between $F_1$ and $F_2$: $-F_1\cos\theta_\ell+F_2/2\sin\theta_\ell=0$. A summary of the boundary-value problem is shown in Fig.~\ref{fig:FigS2}. The solution of the problem is obtained numerically using a shooting method. The shooting method is implemented using Matlab (Matlab 2018b, Mathworks) and the function \textit{ode45} to compute the solution of Eq.~\ref{Eq:elastica} expressed as a first order differential equation with 4 variables $Y=[\theta(S) , \theta'(S) , x_1(S) , x_2(S)]$ and $Y'=[Y(2) , \left({1-\left({\frac{nh}{2}Y(2)}\right)^2}\right)^2\times|\mathbf{F}|\left(\sin\theta_\ell\cos Y(1)+\cos\theta_\ell\sin Y(1)\right) , \cos(Y(1)) , \sin(Y(1))]$. For a given deflection and starting from $S=\ell$, the parameters $(\theta_\ell,\ell,|\mathbf{F}|)$ are varied using \textit{fsolve} until the boundary conditions in $S=0$ are verified.

\section{Internal stress in the backbone solution}

\noindent In this section, we provide a detailed justification for the elastic backbone (friction-less case), and we identify the microscopic stress, based on which, the expression for the power dissipated by friction is also obtained to first order in the friction coefficient. \\

\noindent \textbf{Microscopic equations of equilibrium:}
We start with a detailed analysis of the elastic backbone model. We use two coordinate systems, $(S, y)$ and $(\tilde{S}, y)$, where
$\tilde{S}$ is the (Lagrangian) arc length along a plate, and $S$ is the (non-Lagrangian) arclength of the projection of the current point onto the centerline in actual configuration. Each layer in the stack is in equilibrium. As a result, the shear and normal
forces of the individual layers, denoted as $\tilde{T}$ and $\tilde{N}$
respectively, and their internal moment $\tilde{M}$ must satisfy the Kirchhoff
equations for the equilibrium of elastic rods everywhere,
\begin{equation}
  \begin{array}{cllll}
    \frac{\partial \tilde{M}}{\partial \tilde{S}} (\tilde{S}, y) + \tilde{T}
    (\tilde{S}, y) &  &  & = & 0\\
    \frac{\partial \tilde{N}}{\partial \tilde{S}} (\tilde{S}, y) -
    \tilde{\kappa} (\tilde{S}, y)  \tilde{T} (\tilde{S}, y) &  &  & = & 0\\
    \frac{\partial \tilde{T}}{\partial \tilde{S}} (\tilde{S}, y) +
    \tilde{\kappa} (\tilde{S}, y)  \tilde{N} (\tilde{S}, y) & + &
    \tilde{p}_{\text{n}} (\tilde{S}, y) & = & 0
  \end{array} \label{eq:localKirchoffEquil}
\end{equation}
Note that we follow the standard (but potentially confusing) convention whereby the normal force $\tilde{N}$ is along the tangent to the layer and the shear force $\tilde{T}$ is perpendicular to the layer.
In this frictionless model, the loading applied on each layer is the
\textit{net} transverse force $\tilde{p}_{\text{n}} \mathd \tilde{S}$;
\textit{i.e.}, the balance of transverse forces applied by the adjacent layers.

\noindent \textbf{Microscopic shear force:} Along a given layer, the transverse coordinate $y$ is constant. The shear force $\tilde{T} (\tilde{S}, y)$ can be found by combining the balance of moments in Eq.~(\ref{eq:localKirchoffEquil}), with the constitutive law $\tilde{M} = B_1 \tilde{\kappa}$ and using the expressions of the plate curvature $\tilde{\kappa} (\tilde{S}, y)$ in terms of the centerline curvature $\kappa (S)$ from Eq.~(\ref{Eq:KappaTilde}), as well as $\mathd \tilde{S}$ from Eq.~(\ref{Eq:layer_metric}):
\begin{equation}
  \begin{array}{lll}
    \tilde{T} & = & - \frac{\partial \tilde{M}}{\partial \tilde{S}}\\
    & = & - \frac{\partial \tilde{M}}{\partial S}  \frac{1}{1 - \kappa y}\\
    & = & - \frac{B_1}{1 - \kappa y}  \frac{\partial \tilde{\kappa}}{\partial
    S}\\
    & = & - \frac{B_1}{1 - \kappa y}  \frac{\partial}{\partial S} \left(
    \frac{\kappa}{1 - \kappa y} \right)\\
    & = & - \frac{B_1}{1 - \kappa y}  \frac{\partial \kappa}{\partial S} 
    \left( \frac{1}{1 - \kappa y} + \frac{\kappa y}{(1 - \kappa y)^2}
    \right)\\
    & = & - \frac{B_1}{(1 - \kappa y)^3}  \frac{\mathd \kappa}{\mathd S} .
  \end{array} \label{Eq:T}
\end{equation}

\noindent \textbf{Microscopic normal force:} In view of the longitudinal balance of forces in
Eq.~(\ref{eq:localKirchoffEquil}), the normal force satisfies:
\[ \begin{array}{lll}
     \frac{\partial \tilde{N}}{\partial S} (S, y) & = & \frac{\partial
     \tilde{N}}{\partial \tilde{S}}  (1 - \kappa (S) y)\\
     & = & \tilde{\kappa}  \tilde{T}  (1 - \kappa (S) y)\\
     & = & \kappa \tilde{T}\\
     & = & - \frac{B_1 \kappa}{(1 - \kappa y)^3}  \frac{\mathd \kappa}{\mathd
     S}\\
     & = & - \frac{B_1}{y^2}  \frac{\kappa y}{(1 - \kappa y)^3}  \left( y
     \frac{\mathd \kappa}{\mathd S} \right)\\
     & = & \frac{B_1}{y^2}  \frac{\partial \psi (\kappa (S) y)}{\partial S}\\
     & = & \frac{\partial}{\partial S}  \left[ \frac{B_1}{y^2} \psi (\kappa
     (S) y) \right],
   \end{array} \]
where we have introduced the auxiliary function $\psi (k) = - \frac{k^2}{2 (1
- k)^2}$, with deriviative $\psi' (k) = - \frac{k}{(1 - k)^3}$.

By integrating $\partial \tilde{N} / \partial S$ we find the normal force as
\[ \begin{array}{lll}
     \tilde{N} (S, y) & = & \frac{B_1}{y^2} \psi (\kappa y) + \tmop{Cte} (y)
     \nonumber\\
     & = & - \frac{B_1 \kappa^2}{2 (1 - \kappa y)^2} + \tmop{Cte} (y)
   \end{array} \]
The integration we just did cannot be carried out across the points of
discontinuity. As a result, the constant of integration $\tmop{Cte} (y)$ may
be different in each of the regions separated by the point-like forces. In
particular, in the half-domain $0 \leqslant S \leqslant L$, there is {\textit{a
priori}} one function $\tmop{Cte} (y)$ in the interval $0 \leqslant S \leqslant \ell$ and a different function $\tmop{Cte} (y)$ in the interval
$\ell \leqslant S \leqslant L$:
\begin{itemize}
  \item Beyond the supports ($\leqslant S \leqslant \ell$), the plates are underformed so $\kappa = 0$, and
  $\tilde{N} (S, y) = \tmop{Cte} (y)$. The free-boundary condition at $S =
  L$ sets $\tmop{Cte} (y) = 0$ and, hence, $\tilde{N} (S, y) = 0$;
  
  \item At the roller ($S = \ell$), the applied force is purely transverse, implying that the normal force $\tilde{N}$ is actually continuous,
  $\llbracket \tilde{N} \rrbracket_{\ell} = \tilde{N} (\ell^+, y) - \tilde{N}
  (\ell^-, y) = - \tilde{N} (\ell^-, y) = 0$. Since the moment is zero at $S =
  \ell^-$, so is the curvature, $\kappa (\ell^-) = 0$; then $\tilde{N}
  (\ell^-, y) = 0$ implies $\tmop{Cte} (y)=0$.
\end{itemize}
We have just shown that the quantity $\tmop{Cte} (y)$ is zero everywhere; \textit{i.e.},
\[ \tilde{N} (S, y) = - \frac{B_1}{2}  \left( \frac{\kappa}{1 - \kappa y}
   \right)^2 . \]

\noindent \textbf{Normal stress -- regular part:} Let us now evaluate the normal forces applied from the neighbors to a given
layer $\tilde{p}_{\text{n}} \mathd \tilde{S}$ for $S < \ell$, for which we will use transverse equilibrium,
\[ \begin{array}{lll}
     \tilde{p}_{\text{n}} \mathd \tilde{S} & = & \tilde{p}_{\text{n}} (1 -
     \kappa y) \mathd S\\
     & = & - \left( \frac{\partial \tilde{T}}{\partial \tilde{S}} +
     \tilde{\kappa}  \tilde{N} \right)  (1 - \kappa y) \mathd S\\
     & = & - \left( \frac{\partial \tilde{T}}{\partial S} + \kappa \tilde{N}
     \right) \mathd S \nonumber\\
     & = & B_1  \left( \frac{\partial}{\partial S}  \left( \frac{1}{(1 -
     \kappa y)^3}  \frac{\mathd \kappa}{\mathd S} \right) + \kappa \frac{1}{2}
     \left( \frac{\kappa}{1 - y \kappa} \right)^2 \right) \mathd S\\
     & = & B_1  \left( \frac{1}{2}  \frac{\kappa^3}{(1 - y \kappa)^2} +
     \frac{3 y}{(1 - \kappa y)^4} \left( \frac{\mathd \kappa}{\mathd S}
     \right)^2 + \frac{1}{(1 - \kappa y)^3}  \frac{\mathd^2 \kappa}{\mathd
     S^2} \right) \mathd S.
   \end{array} \]
   
In the main text, we have defined $\Sigma (S, y)$ as the normal stress at the
plate-plate interfaces. The normal force applied by the plate above
(respectively, below) the plate having mean coordinate $y$, over an interface
element with length $\mathd \tilde{S}$, is therefore $- \Sigma (S, y + h / 2)
\mathd \tilde{S}$ (respectively, $+ \Sigma (S, y - h / 2) \mathd \tilde{S}$).
The net force experienced by the plate from the adjacent plates is therefore
$\tilde{p}_{\text{n}} \mathd \tilde{S} = h \frac{\partial (\Sigma \mathd
\tilde{S})}{\partial y}$, which we can rewrite as
\begin{equation}
  h \frac{\partial (\Sigma (1 - \kappa y))}{\partial y} = \tilde{p}_{\text{n}}
  (1 - \kappa y) .
  \label{eq:netTransverseForceOnLayer}
\end{equation}
This equation can be integrated with respect to $y$, using the free boundary
conditions at top and bottom of the stack $\Sigma (S, \pm \frac{nh}{2}) = 0$;
this yields the normal stress in the elastic backbone solution as
\begin{equation}
  \Sigma (S, y) = \frac{1}{h (1 - \kappa (S) y)}  \int_{- \frac{nh}{2}}^y B_1 
  \left( \frac{1}{2}  \frac{\kappa^3 (S)}{(1 - \kappa (S)  \tilde{y})^2} +
  \frac{3 \tilde{y}}{(1 - \kappa (S)  \tilde{y})^4} \left( \frac{\mathd
  \kappa}{\mathd S} \right)^2 + \frac{1}{(1 - \kappa (S)  \tilde{y})^3} 
  \frac{\mathd^2 \kappa}{\mathd S^2} \right) \mathd \tilde{y} .
  \label{eq:normalstress}
\end{equation}
One can check that the stress-free condition at the top of the stack $\Sigma
(S, y = \frac{nh}{2}) = 0$ is automatically satisfied, even if it has not been
enforced.

\begin{figure}
\includegraphics[width=0.7\columnwidth]{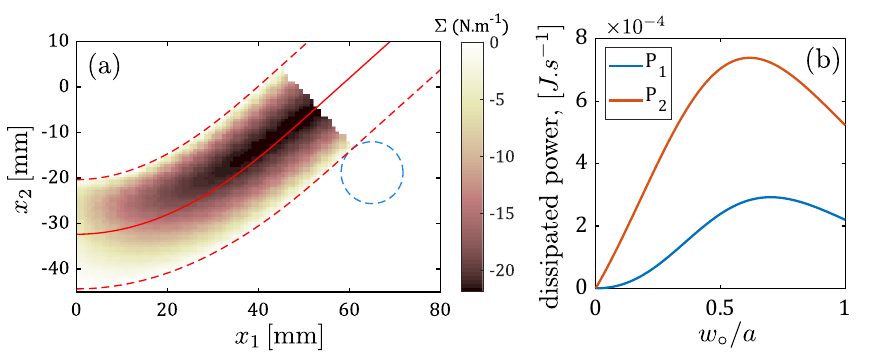}
\caption{\label{fig:FigS3} (a)~Normal stress in the stack $\Sigma(S,y)$ at indentation level $w_0/a=0.5$ (b)~Comparison between the power dissipated away from the point of contact ($S<\ell$), $\mathcal P_1$ and at the contact point $\mathcal P_2$. For both (a) and (b) the following parameters has been used: $a=65\,\mathrm{mm}$, $h=0.286\,\mathrm{mm}$, $B_1=1.76.10^{-4}\,\mathrm{N.m^2}$, $b_0=6.8\,\mathrm{mm}$, $n=80$, $\dot{w_\circ}=1\,\mathrm{mm.s^{-1}}$ and $\mu=0.52$. 
}
\end{figure}

The normal stress $\Sigma (S, y)$ is evaluated and represented
in~Fig.~\ref{fig:FigS3}(a). We observe that the pressure is always negative,
meaning that the plates are pressing against each other. \\

\noindent \textbf{Normal stress -- singular contribution at the rollers:} The expression in Eq.~(\ref{eq:normalstress}) for the normal stress is valid away from the points $S \in
\{ - \ell, 0, \ell \}$, where point-like forces are applied. We do not need to
derive the singular normal stress at the point of indentation $S = 0$, since
the sliding velocity of the plates is zero there by symmetry, implying that
there is no frictional dissipation. 

We proceed to derive the singular (Dirac-like) contribution of the internal stress at the roller $S = \ell$. The other roller $S = - \ell$ is treated similarly, by symmetry. At $S = \ell$, the point-like force applied by the roller induces a point-like
net normal force $\tilde{p}_{\text{n}}^D$ applied to each plate, leading to
the following balance of forces and moments,
\begin{equation}
  \llbracket \tilde{T} \rrbracket_{\ell} + \tilde{p}_{\text{n}}^D = 0, \quad
  \llbracket \tilde{N} \rrbracket_{\ell} = 0, \quad \llbracket \tilde{M}
  \rrbracket_{\ell} = 0.
\end{equation}
where $\llbracket f \rrbracket_{\ell} = f (y, \ell^+) - f (y, \ell^-)$ denotes
the discontinuity of a function $f$ across $S = \ell$. The equation $\llbracket \tilde{N} \rrbracket_{\ell} = 0$ has already been
used to determine $\tilde{N}$. The equation $\llbracket \tilde{M} \rrbracket_{\ell} = 0$ has been used to
show that the curvature is continuous across $S = \ell$, which motivates the
boundary condition $\kappa (\ell^-) = \theta' (\ell^-) = 0$ used in the
boundary value problem of the elastic backbone.

We insert the expression of $\tilde{T}$ from Eq.~(\ref{Eq:T}) into the balance
of normal forces $\llbracket \tilde{T} \rrbracket_{\ell} +
\tilde{p}_{\text{n}}^D = 0$; noting that $\frac{\mathd \kappa}{\mathd S}
(\ell^+) = 0$, as the plates remain straight for $S > \ell$, we obtain:
\begin{equation}
  B_1  \frac{\mathd \kappa}{\mathd S} (\ell^-) + \tilde{p}_{\text{n}}^D = 0
\end{equation}
By the same argument as earlier, the Dirac-like contribution to the normal
stress $\Sigma^D (y)$ at $S = \ell$ satisfies the differential equation
\[ h \frac{\partial \Sigma^D}{\partial y} = \tilde{p}_{\text{n}}^D \]
as well as the boundary conditions
\[ \Sigma^D \left( - \frac{nh}{2} \right) = - |\mathbf{F}|
   \quad{2 \tmop{em}} \Sigma^D \left( + \frac{nh}{2} \right) = 0 \]
where $\mathbf{F}$ is the point-like force applied by the roller from below. The solution is found by integration as
\begin{equation}
  \Sigma^D (y) = - \frac{1}{h}  \int_y^{+ \frac{nh}{2}} \tilde{p}_{\text{n}}^D
  \mathd \tilde{y} = \frac{1}{h} B_1  \frac{\mathd \kappa}{\mathd S} (\ell^-) 
  \left( \frac{nh}{2} - y \right) .
\end{equation}
The constant of integration warrants the equilibrium on the topmost
interface. The equilibrium of the bottommost interface yields
\begin{equation}
  nB_1  \frac{\mathd \kappa}{\mathd S} (\ell^-) = - |\mathbf{F}| .
\end{equation}
This equation can be interpreted as a balance of transverse force for the
one-dimensional model across the singularity (details not shown). In view of this, the singular contribution to the transverse stress at $S =
\ell$ can be rewritten as
\begin{equation}
  \Sigma^D (y) = - |\mathbf{F}|  \left( \frac{1}{2} - \frac{y}{nh} \right) .
  \label{eq:SigmaD}
\end{equation}

\noindent \textbf{Sliding velocity:} We define the sliding displacement as
\[ u (y, S) = \tilde{S} (y, S) - S, \]
which satisfies
\begin{equation}
  \frac{\partial u}{\partial S} = \frac{\partial \tilde{S}}{\partial S} - 1 =
  - \kappa (S) y,
\end{equation}
and, hence,
\begin{equation}
  u (y, S) = - y \theta (S) .
\end{equation}
The relative displacement at an interface therefore reads
\begin{equation}
  \delta (y, S) = \tilde{S} \left( y + \frac{h}{2}, S \right) - \tilde{S}
  \left( y - \frac{h}{2}, S \right) = u \left( y + \frac{h}{2}, S \right) - u
  \left( y - \frac{h}{2}, S \right) = h \frac{\partial u}{\partial y} (u, S) =
  - h \theta (S) .
\end{equation}
The time derivative of this expression yields the relative sliding velocity:
\begin{equation}
  \dot{\delta} (y, S) = - h \dot{\theta} (S) .
\end{equation}

\noindent \textbf{Perturbative expression for the power dissipated by frictional forces:} Throughout our study, friction is treated as a perturbation; we assume that friction does not significantly affect the microscopic stress in the stack, nor the sliding velocities at the interfaces. We use an Amontons-Coulomb friction law
between the plates, which yields the tangent stress at the interfaces between plates as $\mu \Sigma$, where $\mu$ is the friction coefficient. There are also singular (Dirac-like) tangent force at the points with coordinates $S \in \{ - \ell, 0, \ell \}$ on each of the interfaces; that corresponding to $S = \ell$ reads $\mu \Sigma^D$. Now, we seek to compute the power dissipated by friction in the entire stack $\mathcal{P}_\mu$, for which we have to first consider two separate contributions, $\mathcal{P}_1$ and $\mathcal{P}_2$, which are detailed next.

The power $\mathcal{P}_1$ dissipated by friction away from the singular points
$S \in \{ - \ell, 0, \ell \}$ is the integral over all plate-plate interfaces
of $\mu \Sigma$ times the sliding velocity $\dot{\delta}$:
\begin{equation} \mathcal{P}_1 = \mu \int_{- L}^L \mathd S \int_{- \frac{nh}{2}}^{+
   \frac{nh}{2}} h | \dot{\theta} (S) |  (- \Sigma (S, y))  \frac{\mathd y}{h}
    \label{eq:P1Raw}
\end{equation}
where we assume that $\Sigma (S, y) < 0$ everywhere, as we checked. In this
expression, there is an integral $\int \cdots \mathd S$ along the interfaces,
and an integral $\int \cdots \frac{\mathd y}{h}$ which serves as a continuous
approximation to the sum over all interfaces. Performing the integration in Eq.~(\ref{eq:P1Raw}) with respect to the transverse variable $y$, by (i) omitting the regions
$| S | > \ell$ where the normal stress is zero, (ii) identifying the quantity $Q
(S) = \int_{- \frac{nh}{2}}^{+ \frac{nh}{2}} (- \Sigma (S, y)) \mathd y$ defined in the main text, and (ii) limiting the integration to the domain $S \geqslant 0$ by symmetry, yields
\[ \mathcal{P}_1 = 2 \mu \int_0^{\ell} Q (S) | \dot{\theta} (S) | \mathd S. \]
The power dissipated by friction caused by the Dirac-like contribution at $S =
0$ is zero, since the sliding velocity is zero there, $\dot{\delta} (0, S) =
0$.

The power $\mathcal{P}_2$ dissipated by friction caused by the Dirac-like contribution at $S = \ell$ writes, by a similar argument, as
\[ \mathcal{P}_2 = \mu \int_{- \frac{nh}{2}}^{+ \frac{nh}{2}} h | \dot{\theta}
   (\ell) |  (- \Sigma^D (y))  \frac{\mathd y}{h} . \]
Using Eq.~(\ref{eq:SigmaD}), we can calculate the intregral as
\begin{equation} \mathcal{P}_2 = \mu hn | \dot{\theta} (\ell) |    \frac{|\mathbf{F}|}{2} .
  \label{eq:F2_explicit}
\end{equation}

For a given value for the friction coefficient $\mu$, $\mathcal{P}_1$ appears
to be somewhat smaller than (but still comparable to) $\mathcal{P}_2$, see
Fig.~\ref{fig:FigS3}(b). Finally, the power dissipated by friction in the entire stack is
\begin{equation}
    \mathcal{P}_{\mu} =\mathcal{P}_1 + 2\mathcal{P}_2, 
    \label{eq:PmuAsASum}
\end{equation}
where the factor 2 is because there are two rollers. This is the expression used in the main text.

\section{The case of small deflections}

In this section, we consider the limit where the deflection is small, $w_{\circ} / a \ll 1$, and the stack is overall slender, $nh / a \ll 1$.

We start with a linear analysis of the backbone solution. In this limit, the stack's centerline
can be analyzed using the approximations $\cos \theta = 1$, $\sin \theta = \theta$ and $\ell = a$. The linearized equilibrium equations in the transverse
direction write
\[ nB_1  \frac{\mathd^2 \theta}{\mathd S^2} + \frac{F_2}{2} = 0 \quad
   \frac{\mathd x_2}{\mathd S} = \theta, \]
with the boundary conditions: $x_2 (0) = -w_{\circ}$, $x_2 (a) = 0$, $\theta
(0) = 0$ and $\frac{\mathd \theta}{\mathd S} (a) = 0$. This equivalent, linear
beam problem, with a bending stiffness $nB_1$, is solved as
\begin{equation}
  \begin{array}{rll}
    \theta_{\text{bb}} (S) & = & \frac{F_2 a^2}{2 nB_1}  \left( \frac{S}{a} -
    \frac{1}{2} \left( \frac{S}{a} \right)^2 \right)\\
    x_{2, \text{bb}} (S) & = & \frac{F_2 a^3}{4 nB_1}  \left( - \left( 1 -
    \left( \frac{S}{a} \right)^2 \right) + \frac{1}{3}  \left( 1 - \left(
    \frac{S}{a} \right)^3 \right) \right)
  \end{array} \label{eq:linearSolution}
\end{equation}

This result implies the linear indentation law $F_{2, \text{bb}} = \frac{6 nB_1}{a^3}
w_{\circ}$. Consequently, the elastic energy of the stack writes as $\mathcal{E} (w_{\circ})
= \frac{{3} nB_1}{a^3} w_{\circ}^2$, from which  we recover the incremental rigidity $K_{\text{bb}} (w_{\circ}) = \frac{a^3}{6}  \frac{\mathrm{d} F_{2, \text{bb}}}{\mathrm{d} w_{\circ}} = nB_1$ announced in the main text.

In the limit $nh / a \ll 1$, all the plates have the same shape, given by the
centerline plus a constant vertical offset; {\textit{i.e.}},
$\tilde{\mathbf{x}} (S, y) = S\mathbf{e}_1 + \left( x_{2, \text{bb}} (S) +
y \right) \mathbf{e}_2$, with $\tilde{S} = S$ in this linearized setting.
Inserting this result into the local equations of equilibrium for the individual
layers, Eq.~(\ref{eq:localKirchoffEquil}), one finds $\tilde{p}_{\text{n}} = 0$, meaning that 
away from the indentation point and from the rollers, $S \not\in \{ - \ell, 0,
\ell \}$, each layer is in equilibrium without any applied force. In view of
Eq.~(\ref{eq:netTransverseForceOnLayer}), the normal stress $\Sigma (S, y)$ is
independent of $y$. With the stress-free boundary conditions on the uppermost
and lowermost faces, we have that
\[ \Sigma (S, y) = 0, \]
in the linear regime. In view of Eq.~(\ref{eq:P1Raw}), this result implies
$\mathcal{P}_1 = 0$; {\textit{i.e.}}, for small friction and small deflection,
dissipation occurs dominantly at the rollers, at $S = \pm \ell$.

Using Eqs.~(\ref{eq:F2_explicit}--\ref{eq:PmuAsASum}), we find that the power
dissipated by friction can be estimated based on the elastic backbone solution
as \ $\mathcal{P}_{\mu} = \frac{\mu nh | \mathbf{F} |}{2} |
\dot{\theta}_{\text{bb}} (a) |$. In the linear regime, $\mathbf{F}$ can be
approximated as $F_2/2 \mathbf{e}_2$; {\textit{i.e.}},
    \begin{equation}
    \mathcal{P}_{\mu} = \frac{\mu nhF_2}{2} | \dot{\theta}_{\text{bb}} (a) | 
    \label{eq:SI:Pmuintermediate}
    \end{equation}

We know that $F_{2, \text{bb}} = \frac{6 nB_1}{a^3} w_{\circ}$, and
Eq.~(\ref{eq:linearSolution}) shows that $\theta_{\text{bb}} (a) = \frac{F_{2,
\text{bb}} a^2}{4 nB_1} = \frac{3}{2 a} w_{\circ}$. Inserting this result into the
right-hand side of Eq.~(\ref{eq:SI:Pmuintermediate}), we find
\[ \mathcal{P}_{\mu} = \frac{9}{2}  \frac{\mu hn^{2} B_1}{a^4}
   w_{\circ} | \dot{w_{\circ}} | . \]

The expression of the indentation force $F_2^{\pm} (w_{\circ})$ can now be derived using Eq~(4) from the main text, while taking into account the frictional dissipated energy, as 
\begin{equation}
  F_2^{\pm} = \frac{6}{a^2} nB_1  \left( 1 \pm \frac{3}{4} \mu n \frac{h}{a}
  \right) \frac{w_{\circ}}{a}, \label{eq:F2-loading-unloading}
\end{equation}
with the $+$ sign for loading and $-$ for unloading. Therefore, our linear theory predicts an apparent, incremental (scaled) stiffness
\begin{equation} K_{\mathrm{lin}} = nB_1  \left( 1 \pm \frac{3}{4} \mu n \frac{h}{a}
   \right), 
   \label{eq:KLin}
\end{equation}
where friction causes an apparent stiffening upon loading and an apparent softening upon unloading. 

The expression of $F_2$ for loading and unloading in Eq.~(\ref{eq:F2-loading-unloading}) yields the
following prediction for the energy dissipated during one cycle:
\[ \begin{array}{lll}
     D_{\mathrm{lin}} & = & \int_0^{w_{\circ}^{max}} F_2^+ \mathd w_{\circ} -
     \int_0^{w_{\circ}^{max}} F_2^- \mathd w_{\circ}  \nonumber\\
     & = & \frac{9 \mu B_1 h}{2 a^2}  \left( n
     \frac{w_{\circ}^{\mathrm{max}}}{a} \right)^2 ,
   \end{array} \]
which is plotted as the dashed line in Fig.~3(b) of the main text.

\section{Measure of the effective friction coefficient $\mu$}

\begin{figure}[b]
\includegraphics[width=0.7\columnwidth]{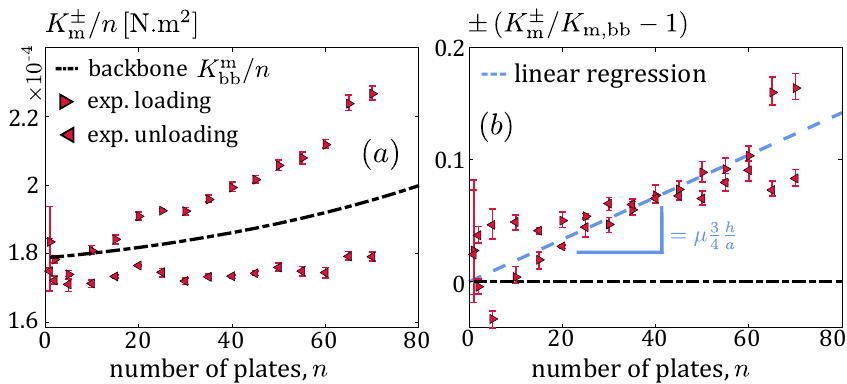}
\caption{\label{fig:FigS4} Extraction of the friction coefficient from experimental data. (a)~Scaled maximum stiffness $K_\mathrm{m}^{\pm}/n$ as a function of $n$, for loading ($\rhd$) and unloading ($\lhd$). The dashed-dotted curve is the prediction from the elastic backbone solution. (b)~Increase in apparent rigidity due to friction: the slope from the linear regression versus $n$ (dashed blue line) provides a measure of the friction coefficient $\mu$.
}\end{figure}

In Eq.~(\ref{eq:KLin}) above, we wrote the bending rigidity of the stack, $K_\mathrm{lin}$, for the linear case. We extrapolate this prediction to the frictional nonlinear regime by applying the same corrective factor $\left(1\pm \frac{3}{4}\mu n \frac{h}{a}\right)$ to the friction-less, geometrically nonlinear prediction for the maximum incremental stiffness (elastic backbone) $K_\mathrm{m,bb} = \max_{w_\circ}(K_\mathrm{bb})$ where 
$K_\mathrm{bb}(w_\circ)=\frac{a^3}{6}\,\mathrm{d}F_{2,\mathrm{bb}}/\mathrm{d}w_\circ$ (see Fig.~\ref{fig:FigS4}a):
\begin{equation}
 K_\mathrm{m}^{\pm} = K_\mathrm{m,bb}\cdot\left(1\pm \frac{3}{4}\mu n\frac{h}{a}\right).
\end{equation}
The increase in apparent rigidity due to friction $ \pm(K_\mathrm{m}^{\pm} / K_\mathrm{m,bb} - 1 ) = \frac{3}{4}\mu n\frac{h}{a}$ should therefore depend linearly on $n$, see  Fig.~\ref{fig:FigS4}(b).
This prediction is indeed verified by the experimental data for both loading and unloading, which enables us to extract the friction coefficient as $\mu=0.52\pm0.03$.

